\documentstyle[12pt]{article}
\def\beqa{\begin{eqnarray}}
\def\eeqa{\end{eqnarray}}
\def\beq{\begin{equation}}
\def\eeq{\end{equation}}

\def\ie{{\it i.e. }}
\def\eg{{\it e.g. }}

\def\pr{{\it Phys. Rev.}\ }
\def\prl{{\it Phys. Rev. Lett.}\ }
\def\pl{{\it Phys. Lett.}\ }
\def\np{{\it Nucl. Phys.}\ }

\def\ijmp{{\it Int. Journ. Mod. Phys.}\ }

\def\cqg{{\it Class. Quantum Grav.}\ }

\def\grg{{\it Gen. Relativ. Grav.}\ }

\def\rmp{{\it Rev. Mod. Phys.}\ }

\begin{document}
\def\bib#1{[{\ref{#1}}]}
\begin{titlepage}
    \title{Nonminimal Derivative Coupling and  \\
           the Recovering of Cosmological Constant}

\author{S. Capozziello\thanks{E-mail:capozziello@vaxsa.csied.unisa.it}~ and
G. Lambiase\thanks{E-mail:lambiase@vaxsa.csied.unisa.it} \\
{\em Dipartimento di Scienze Fisiche ``E. R. Caianiello'',} \\
{\em Universit\`{a} di Salerno, I-84081 Baronissi, Salerno, Italy.} \\
{\em Istituto Nazionale di Fisica Nucleare, Sezione di Napoli, Italy} \\
}
\date{\today}
\maketitle
\begin{abstract}
We show that the existence of the cosmological constant
can be connected to a nonminimal
derivative coupling, in the action of gravity, between the geometry
and the kinetic part of a given scalar field
 without introducing any effective potential of scalar fields.
Exact solutions are given.
\end{abstract}
\thispagestyle{empty}

\vspace{20. mm}


\vfill

\end{titlepage}

\section{\normalsize\bf Introduction}

Including nonlinear terms of the various curvature tensors
(Riemann, Ricci, Weyl) and nonminimally coupled
terms in the effective action of gravity has become, recently, a very
common trend from quantum field theory side and cosmology
\cite{wands}.
The basic motivation for studying such
theories comes from the fact that they provide a possible
approach to quantum gravity from a perturbative point of view.
Furthermore they occur as low--energy limit of
several unification scheme as, \eg superstring theory \cite{green}.

Modern cosmology, starting from the pioneering works by Starobinsky
\cite{starobinski}, has found into them a fruitful arena for trying
to solve the several shortcomings of standard
cosmological model as initial singularity, flatness, horizon problems
and so on, in the framework of the inflationary paradigm.

In fact, nonminimal coupling between  scalar field(s) and 
geometry and higher order
terms in the curvature invariants naturally give rise to inflationary
solutions which, in various senses,  improve the early inflationary models
(see for example \cite{la},\cite{aclo}).

Another important question  connected to such theories is a dynamical
 determination of cosmological constant
which could furnish the gravity vacuum state  \cite{weinberg}
and could contribute to solve the dark matter problem:
in fact, the presence of a cosmological constant gives rise to viable
models for large scale structure \cite{carrol} as recent observations are
confirming \cite{lensing}.

Besides, the exact determination of cosmological constant could account
for the fate of the whole Universe considering
the so called no--hair conjecture \cite{hoyle}. In any
case, we need a time variation of cosmological constant to satisfy
issues as successful inflationary
models, the agreement with large--scale structure observations, 
and to obtain a de
Sitter stage in the future, if a {\it remnant} of cosmological constant
is present into the overall dynamics.

In other words, the cosmological constant should have acquired
high values at early
times (de Sitter stage), should have undergone a phase transition with
a graceful exit (to recover the observed dust dominated Friedman stage)
and should result in a remnant in the future.

Considering the wide variety of extended gravity theories which can give
rise to de Sitter stages (\ie where cosmological constant leads dynamics)
a main question is to
recover classes of gravitational theories which ``naturally" give rise
to cosmological
constant without putting it ``by hand" or considering special initial
data \cite{lambda}.
Furthermore, is cosmological constant related only with
the presence of an effective potential of a scalar field?
Can it be recovered also by introducing nonminimal couplings
between geometry and scalar fields without introducing any sort of potential?

We have to stress the fact that also considering
pure higher order theories as $f(R)= R+\alpha R^2+\ldots$,
by a conformal transformation higher order terms give rive
effective potential. In general, these theories have de Sitter
stages \cite{barrow},\cite{kluske}.

In this paper we want to show that it is possible to recover the
de Sitter behaviour, and
then the cosmological constant  by introducing nonminimal
derivative couplings between the geometry and the scalar field.

These kind of couplings naturally arise in Kaluza--Klein
and superstring theories
\cite{linde} while they must be included in the matter Lagrangian of
quantum field theories in curved spacetimes involving
scalar fields and a multiloop expansion \cite{donoghue}.

From a general point of view the effective Lagrangian of
quantum gravity, considering
the  expansion of curvature invariants and the matter fields,
can be written as \cite{buchbinder}
\beq
\label{Leff}
{\cal L}_{effective}={\cal L}_{g}+{\cal L}_{m}\,{,}
\eeq
where
\beq
\label{Lg}
{\cal L}_{g}=\sqrt{-g}\left\{\Lambda
-\frac{R}{2}+a_1R^2+a_2R_{\mu\nu}R^{\mu\nu}+{\cal O}(R^3)\right\}\,{,}
\eeq
and
\beq
\label{Lm}
{\cal L}_{m}=\sqrt{-g}\left\{\frac{1}{2}
(\partial_{\mu}\phi\partial^{\mu}\phi-m^2\phi^2)+
\right.
\eeq
$$ \left.
+d_1R^{\mu\nu}\partial_{\mu}\phi\partial_{\nu}\phi
+R(d_2\partial_{\mu}\phi\partial^{\mu}\phi
+d_3m^2\phi^2)+\ldots\right\}\,{.}
$$
We are assuming $8\pi G=1$;  $a_i$ and $d_i$ are coupling constants which
scale as powers of the mass. Here the gravitational Lagrangian
has been ordered in
a derivative expansion of the metric
with $\Lambda$ being of order $\partial^0$, $R$ of the order
$\partial^2$, $R^2$ and $R_{\mu\nu}R^{\mu\nu}$
of order $\partial^4$, and so on.

In four dimensions, we do not need to include terms as
$R_{\alpha\beta\mu\nu}R^{\alpha\beta\mu\nu}$ in the action
since, by the Gauss--Bonnet
theorem, it is possible to express these terms as
$R^2$ and $R_{\mu\nu}R^{\mu\nu}$.
Several papers have been devoted to the $R^2$ and
scalar field cosmologies but not so
much have pointed out the relevance of
derivative coupling in order to recover the
cosmological constant. In \cite{amendola}, for example,
a systematic study of phase
space and inflationary attractors is done for this kind of cosmologies.
However, in our
knowledge, it has never been stressed how it is possible to recover
``exactly" de Sitter
behaviours and cosmological constant starting from them.

This issue takes relevance in the debate of how to recover the
vacuum state in general relativity. If the de Sitter stage
is obtained without considering effective scalar fields
potential as $V(\phi )\simeq (\lambda/n)\phi^n$ or
$V(\phi )\simeq m^2\phi^2$ (or the conformal transformed field potential
starting from theories as $R^2$), it means,
in our opinion, that these terms are not so essential for recovering
the cosmological constant (in this case from the matter field side).

In Sec.2, we consider an action where the nonminimal
derivative coupling is introduced
in a simple way. We show that de Sitter solutions exists. Sec.3
is devoted to the
study of an action in which nonminimal coupling is introduced
for the field $\phi$ and
for its covariant derivative $\phi_{\mu}\equiv \nabla_{\mu}\phi$.
 A general discussion on how it is
possible to construct an effective cosmological constant and how to recover
asymptotically a de Sitter behaviour is done in Sec.4.
We follow the method outlined
in \cite{lambda},\cite{lambdagen}. Conclusions are drawn in Sec.5.

\section{\normalsize\bf Minimal Coupling
with  Nonminimal Derivative Coupling}

Let us start our considerations from the action
\beq
\label{action}
{\cal A}=\int d^4x \sqrt{-g}\left[-\frac{R}{2}+
\frac{1}{2}g^{\mu\nu}\partial_{\mu}\phi
\partial_{\nu}\phi+\zeta R^{\mu\nu}\partial_{\mu}\phi\partial_{\nu}\phi+\xi R
g^{\mu\nu}\partial_{\mu}\phi\partial_{\nu}\phi\right]\,{.}
\eeq
We have not introduced any effective scalar field potential.
In a Friedman--Robertson--Walker (FRW) metric, the action (\ref{action})
reduces to the form
\beq
\label{action1}
{\cal A}=2\pi^2\int dt a^3\left\{\left
[\left(\frac{\ddot{a}}{a}\right)+
\left(\frac{\dot{a}}{a}\right)^2+\frac{k}{a^2}\right](3-6\xi\dot{\phi}^2)+
\frac{\dot{\phi}^2}{2}-3\zeta \left(\frac{\ddot{a}}{a}\right)\dot{\phi}^2\right\} \,{.}
\eeq
Integrating by parts and eliminating the boundary terms,
we get the pointlike Lagrangian
\beq
\label{Lpoint} {\cal L}=3a\dot{a}^2(1+\eta\psi^2)+6\chi
a^2\dot{a}\psi\dot{\psi}-\frac{1}{2}a^3 \psi^2\,{,}
\eeq
where
\beq
\label{cost} \eta
=-2(\xi+\zeta ), \qquad \chi =-(2\xi +\zeta )\,{,} \eeq
and we are considering, for simplicity,  the
spatially flat case ($k=0$). To reduce the degree of the
derivative term of scalar
field, we define the auxiliary field
\beq
\label{psi} \psi =\dot{\phi}\,{,}
\eeq
so that
the Lagrangian (\ref{Lpoint}) assumes the canonical form
${\cal L}={\cal L}(a,\dot{a}, \psi, \dot{\psi})$.
The equations of motion are
\beq
\label{Eq1}
(2\dot{H}+3H^2)(1+\eta\psi^2)+4\eta H\psi\dot{\psi}+2\chi\dot{\psi}^2
+2\chi\psi\ddot{\psi}+\frac{\psi^2}{2}=0\,{,}
\eeq
\beq
\label{Eq2}
6\chi(\dot{H}+3H^2)=6\eta H^2-1\,{,}
\eeq
\beq
\label{Eq3}
3H^2(1+\eta\psi^2)+6\chi
H\psi\dot{\psi}+\frac{\psi^2}{2}=0\,{,}
\eeq
where $H=\dot{a}/a$ is the Hubble
parameter. Immediately we see that the particular solution
\beq
\label{deSit}
\dot{\psi}=0 \to \psi =\psi_0,  \quad H^2=\frac{\Lambda}{3}\,{,}
\eeq
which is de Sitter, exists and
\beq
\label{cost1}
\Lambda =\frac{1}{2(4\xi +\zeta)}, \quad
\psi_0=\frac{1}{\sqrt{\zeta -2\xi}}\,{.}
\eeq
The cosmological behaviour is then given
by
\beq
\label{adeS}
a(t)=a_0\exp \sqrt{\frac{\Lambda}{3}}\, t\,{,}
\eeq
\beq
\label{fdeS}
\phi (t)=\psi_0 t\,{.}
\eeq
This result tells that a cosmological constant can
be constructed by the parameters of nonminimal derivative coupling.
The general solution
of system (\ref{Eq1})--(\ref{Eq3}) is obtained taking
into account (\ref{Eq2}) which can
be recast as
\beq
\label{Hp} \dot{H}=AH^2+B
\eeq
where
\beq
\label{cost2}
A=\frac{\eta
-3\chi}{\chi}, \qquad B=-\frac{1}{6\chi}\,{.}
\eeq
Two interesting sub--cases, due to the
sign of $B/A$, can be discussed: \\
i) $B/A>0$ implies the solution
\beq
\label{Htan}
H(t)=\sqrt{\frac{B}{A}}\tan \sqrt{AB}\, (t-t_0)\,{.}
\eeq
ii) If $B/A<0$, it follows
that $H^2>\vert B/A\vert $ and then
\beq
\label{Hexp+} H=\sqrt{\frac{\vert B\vert}{\vert
A\vert}}\left[\frac{1+\exp 2\sqrt{\vert A\vert \vert B\vert }\, (t-t_0)} {1-\exp
2\sqrt{\vert A\vert \vert B\vert }\, (t-t_0)}\right]\,{,}
\eeq
or
$H^2<\vert B/A\vert $,
so that
\beq
\label{Hexp-}
H=\sqrt{\frac{\vert B\vert}{\vert A\vert}}\left[\frac{\exp
2\sqrt{\vert A\vert \vert B\vert }\, (t-t_0)-1} {\exp
2\sqrt{\vert A\vert \vert B\vert
}\, (t-t_0)+1}\right]\,{.}
\eeq
In both cases we recover asymptotically the solution
(\ref{deSit}) and (\ref{cost1}).

The time evolution of $\psi$ (and then of $\phi$) is
obtained introducing these first
integrals into Eq.(\ref{Eq1}) or Eq. (\ref{Eq3}).
However Eq.(\ref{Eq3}) is the energy
condition $E_{{\cal L}}=0$ (\ie the $(0, 0)$ Einstein equation)
which gives the constraints
on the initial conditions. Integrating Eq.(\ref{Hexp+}) we get
\beq\label{afin+}
a(t)=a_0\left\{\frac{\exp 2\sqrt{\vert A\vert \vert B\vert}\, t}{
\left[\exp 2\sqrt{\vert A\vert \vert B\vert }\, t -1\right]^2}\right\}^{1/2\vert A\vert }\,{.}
\eeq
From Eq.(\ref{Hexp-})
\beq\label{afin-}
a(t)=a_0\left\{\frac{\left[\exp 2\sqrt{\vert A\vert \vert B\vert}\, t +1
\right]^2}
{\exp 2\sqrt{\vert A\vert \vert B\vert}t}\, \right\}^{1/2\vert A\vert }\,{.}
\eeq
Asymptotically we can get increasing and decreasing exponential behaviour.
However, the first case is of physical interest. Eq. (\ref{Htan}) gives an
oscillatory behaviour for the scale factor $a(t)$. All these solutions are
parameterized by the derivative coupling $\zeta$ and $\xi$.

\section{\normalsize\bf Nonminimal Couplings for
Scalar Field and its Derivative}

The generalization of above considerations can be obtained
by introducing a nonminimal
coupling also in the standard part of gravitational Lagrangian.
The effective action becomes
\beq
\label{actionF}
{\cal A}=\int d^4x
\sqrt{-g}\left[F(\phi)R+\frac{1}{2}g^{\mu\nu}\partial_{\mu}\phi
\partial_{\nu}\phi+\zeta R^{\mu\nu}\partial_{\mu}\phi\partial_{\nu}\phi+\xi R
g^{\mu\nu}\partial_{\mu}\phi\partial_{\nu}\phi\right]\,{.}
\eeq
The most general action
should involve terms as
\beq
\label{general}
h(\phi)R^{\mu\nu}\partial_{\mu}\phi\partial_{\nu}\phi\,{,} \quad g(\phi)R
g^{\mu\nu}\partial_{\mu}\phi\partial_{\nu}\phi\,{,}
\eeq
but, for the sake of simplicity,
we restrict to the parameters $\zeta$ and $\xi$.
However, the standard Newtonian coupling
is recovered for $F(\phi)=-1/2$.
Using the above procedure, the pointlike FRW Lagrangian
is
\beq
\label{LpointF}
{\cal L}=6Fa\dot{a}^2+6F^{\prime}a^2\dot{a}\dot{\phi}-6Fka
+\frac{1}{2}a^3\dot{\phi}^2+
\eeq
$$
+6\zeta (a\dot{a}^2\dot{\phi}^2+a^2\dot{a}\dot{\phi}\ddot{\phi})+ 6\xi
(a\dot{a}^2\dot{\phi}^2+2a^2\dot{a}\dot{\phi}\ddot{\phi}-ka\dot{\phi}^2)
\,{.}
$$ where
the prime represents the derivative with respect the scalar filed $\phi$.
In order to
make it canonical, we have to impose
\beq
\label{cond}
\zeta =-2\xi\,{.}
\eeq
Then, Eq.(\ref{LpointF}) reads
\beq
\label{LpointF1} {\cal
L}=6Fa\dot{a}^2+6F^{\prime}a^2\dot{a}\dot{\phi}+\frac{1}{2}a^3\dot{\phi}^2-6\xi
a\dot{a}^2\dot{\phi}^2\,{,}
\eeq
choosing the spatially flat model $k=0$. The
corresponding equations of motion are
\beq
\label{Eq1F}
\ddot{\phi}+3H\dot{\phi}+6(\dot{H}+2H)F^{\prime}-12\xi
(2H\dot{H}\dot{\phi}+3H^3\dot{\phi} +H^2\ddot{\phi})=0\,{,}
\eeq
\beq
\label{Eq2F}
(2\dot{H}+3H^2)F+2\dot{F}H+\ddot{F}-\xi(2\dot{H}+3H^2)\dot{\phi}^2- 4\xi
H\dot{\phi}\ddot{\phi}-\frac{\dot{\phi}^2}{4}=0\,{,}
\eeq
\beq
\label{Eq3F}
H^2F+H\dot{F}+\frac{1}{12}\,\dot{\phi}^2-3\xi H^2\dot{\phi}^2=0\,{.}
\eeq
The form of
the solutions strictly depends on the form of the coupling $F(\phi )$.
The existence of
the de Sitter solution can be imposed so that the form of the coupling is determined as a
function $F(\phi (t))$ (see also \cite{asfree2}).
Assuming, in general,
\beq\label{deSF} a(t)=a_0e^{\lambda t}\,{,}
\eeq
where $\lambda$ is a constant (not necessarily the ``right''
cosmological constant)
we can derive an equation for $F(\phi (t))$ from
(\ref{Eq1F})--(\ref{Eq3F}). With a little algebra, we get
\beq
\label{EqFF}
\ddot{F}+c_1\dot{F}+c_2F=0\,{,}
\eeq
where the constants $c_{1,2}$ are defined as
\beq
\label{c1c2}
c_1=\frac{(12\xi\lambda^2 -5)\lambda}{(12\xi\lambda^2-1)}, \quad
c_2=6\lambda^2\,{.}
\eeq
It follows that the coupling $F(t)$ and the scalar field $\phi(t)$
are
\beq
\label{F(t)}
F(t)=F_1e^{\alpha_1 t}+F_2 e^{\alpha_2 t}\,{,}
\eeq
\beq
\label{phi(t)}
\phi (t)=\phi_0\pm\sigma_1\int\sqrt{b_1e^{\alpha_1 t}+b_2 e^{\alpha_2
t}}\, dt
\eeq
where
\beq
\label{alpha}
\alpha_{1,2}=-\frac{c_1}{2}\pm\sqrt{\left(\frac{c_1}{2}\right)^2-c_2}\,{,}
\eeq
and
\beq
\label{sigma} \sigma_1=\frac{1}{\sqrt{3\xi\lambda^2-1/12}}\,{.}
\eeq $F_{1,2}$
and
$b_{1,2}$ are integration constants. Asymptotically,
we have $F(t)\to F_2e^{\alpha_2 t}$
and then
\beq
\label{F(phi)}
F(\phi )\sim F_0(\phi -\phi_{0})^2\,{,}
\eeq
where $F_0$ depends
on $\lambda$ and $\xi$.
Also in this case the existence of cosmological constant
strictly depends on the derivative coupling while it can be shown
that it does not exist
if only nonderivative nonminimal coupling is present \cite{lambda}.
In other words,
for a pure Brans--Dicke theory it is not possible
to recover a cosmological constant unless
a scalar field potential is introduced by hands.

Finally, it is last term in the Lagrangian (\ref{LpointF1}) which plays the
fundamental role to obtain the de Sitter behaviour.

\section{\normalsize\bf General Considerations and
the Effective Cosmological Constant}

Following \cite{lambda},\cite{kluske},\cite{lambdagen}, \cite{any},
it is possible to show that
for several extended gravity models, a de Sitter behaviour can be recovered.
This means that Wald's proof of no--hair theorem \cite{wald} can be extended
without putting
``by hands" any cosmological constant but recovering it by general
considerations and
defining an ``effective" cosmological constant.

The proof of this statement can be easily sketched following
the arguments in
\cite{lambda} and \cite{lambdagen}. An asymptotic cosmological constant
is recovered any time that the asymptotic conditions
\beq\label{H2>0}
(H-\Lambda_1)(H-\Lambda_2)\geq 0\,{,}
\eeq
\beq
\label{HP<0} \dot{H}\leq 0\,{,}
\eeq
hold. However $\Lambda_{1,2}$ are constants. In our case,
considering Eq.(\ref{Eq3F}),
we can define an effective cosmological constant as
\beq
\label{Laeff}
\Lambda_{eff\,1,2}=\frac{1}{2(F-3\xi\dot{\phi}^2)}\left[-\dot{F}\pm\sqrt{
\dot{F}^2-\frac{\dot{\phi}^2}{3}\, (F-3\xi\dot{\phi}^2)}\right]\,{,}
\eeq so that, Eq.(\ref{Eq3F}) can be recast in the form
\beq
\label{Eq3FH}
(H-\Lambda_{eff,1})(H-\Lambda_{eff,2})=0\,{.}
\eeq
For the  physical consistency of the
problem, it has to be
\beq
\label{delta1} \dot{F}^2-\frac{\dot{\phi}^2}{3}\,
(F-3\xi\dot{\phi}^2)\geq 0\,{,}
\eeq
or simply
\beq
\label{delta2} F-3\xi\dot{\phi}^2< 0\,{.}
\eeq
The asymptotic cosmological constant is recovered if the conditions
\beq
\label{1a}
\frac{\dot{F}}{F-3\xi\dot{\phi}^2}\longrightarrow\Sigma_{0}\,,
\eeq
\beq
\label{2a}
\frac{\dot{\phi}^2}{F-3\xi\dot{\phi}^2}\longrightarrow\Sigma_{1}\,,
\eeq
hold. $\Sigma_{0,1}$ are constants which determine the asymptotic behaviour
of $F$ and $\phi$ (see also \cite{lambdagen}).

Considering Eq.(\ref{Eq2F}), we have
\beq
\label{HP=H}
\dot{H}=-\frac{3}{2}H^2-\frac{H}{F-\xi\dot{\phi}^2}\frac{d}{dt}(F-\xi\dot{\phi}^2)
-\frac{\ddot{F}}{2(F-\xi\dot{\phi}^2)}+\frac{\dot{\phi}^2}{8(F-\xi\dot{\phi}^2)}\,{.}
\eeq
The theorem is easily shown if
\beq
\label{dFt<0}
\frac{d}{dt}(F-\xi\dot{\phi}^2)<0
\eeq
and
\beq
\label{ddF}
\ddot{F}<0\,{,}
\eeq which are very natural conditions which restrict the possible
form of $F$.
Furthermore, it must be $F<0$ in order to recover
attractive gravity (the standard
coupling is for $F\to -1/2$).
 We stress again the relevant role played by the derivative
 coupling in the construction of cosmological constant.

\section{\normalsize\bf Conclusion}

In this paper, we have outlines the role of nonminimal
derivative coupling in recovering
the de Sitter behaviour and then the cosmological constant.
The main point is the fact
that we have not used any effective potential
but the cosmological constant is strictly
related to the derivative coupling and, as it is shown in \cite{lambda},
it cannot be
recovered if the coupling is not derivative. Also by conformal
transformation
\cite{amendola}, it is possible to show that the
right--hand side of Einstein equations
cannot be written as a scalar field energy--momentum tensor
in its proper meaning.

In conclusion, the cosmological constant
can be reconstructed, at least at a classical level, by the kinetic part of
an intervening scalar field ``without" considering scalar field potential.
In this sense, we can deal with a ``dynamical cosmological constant"
which could come out from the interactions between scalar field matter
and geometry.  A further step in this analysis is to consider more generic
derivative couplings and effective potentials in (\ref{actionF}) in order
to see what is the specific role of all ingredients in
the ``construction" of the effective cosmological constant.


\begin{thebibliography}{99}
\bibitem{wands}
T.V. Ruzmaikina and A.A. Ruzmaikin, {\it JETP} {\bf 30} (1970) 372.\\
K.S. Stelle, \grg {\bf 9} (1978) 353.\\
S. Gottl\"ober, H.--J. Schmidt, and A.A. Starobinsky, \cqg {\bf 7} (1990)
893.\\
D. Wands, \cqg {\bf 11} (1994) 269.
\bibitem{green}
M. Green, J. Schwarz, and E. Witten {\it Superstring Theory}
Cambridge Univ. Press, Cambridge 1987.
\bibitem{starobinski}
A.A. Starobinsky, \pl {\bf 91 B} (1980) 99.
\bibitem{la}
D. La and P.J. Steinhardt, \prl {\bf 62} (1989) 376.\\
D. La,  P.J. Steinhardt  and E.W. Bertschinger,
{\it Phys. Lett.} {\bf B 231} (1989) 231.
\bibitem{aclo}
L. Amendola, S. Capozziello, M. Litterio, and F. Occhionero,
\pr {\bf D 45} (1992) 417.
\bibitem{weinberg}
S. Weinberg, \rmp {\bf 61} (1989) 1.
\bibitem{carrol}
S.M. Carrol, W. Press, {\it Ann. Rev. Astron. Astroph.} {\bf 30} (1992) 499.
\bibitem{lensing}
A.A. Starobinsky, in {\it Cosmoparticle Physics 1}, eds. M.Yu. Khlopov
{\it et al.} Edition Frontiers (1996).\\
D. Jain, N. Panchapakesan, S. Mahajan, and V.B. Bhatia, \ijmp {\bf Vol. 13 A}
No. 24 (1998) 4227.
\bibitem{hoyle}
F. Hoyle and J.V. Narlikar, {\it Proc. R. Soc.} {\bf 273A} (1963) 1.
\bibitem{lambda}
S. Capozziello and R. de Ritis, \grg {\bf 29} (1997) 1425.
\bibitem{barrow}
J. Barrow and A.C. Ottewill, {\it J. Phys. A: Math. Gen.} {\bf 16} (1983)
2757.\\
S. Cotsakis and G. Flessas, \pl {\bf B 319} (1993) 69;\\
A.B.Burd and J.D. Barrow,\np {\bf B308} (1988) 929;\\
J.Yokogawa and K.Maeda, \pl {\bf B207} (1988) 31;\\
J.D. Barrow and G. G\"otz, \pl {\bf B 231} (1989) 228.
\bibitem{kluske}
H.--J. Schmidt, \cqg {\bf 7} (1990) 1023;\\
H.--J. Schmidt, \pr {\bf D 54} (1996) 7906.\\
S. Kluske and H.--J. Schmidt, {\it Astron. Nachr.} {\bf 317} (1996) 337;\\
S. Kluske in :{\it New Frontiers in gravitation } ed. G. Sardanashvily.
\bibitem{linde}
A. Linde {\it Particle Physics and Inflationary Cosmology}
Harwood Academic Press, New York 1990.
\bibitem{donoghue}
J.F. Donoghue, \pr {\bf 50D} (1994) 3874.\\
A. Zee, {\it Phys. Rev. Lett.} {\bf 42} (1979) 417 \\
L. Smolin,  {\it Nucl. Phys.}  {\bf B 160} (1979) 253 \\
S. Adler, {\it Phys. Rev. Lett.} {\bf 44} (1980) 1567 \\
N.D. Birrell  and P.C.W. Davies  {\it Quantum Fields in Curved Space} (1982)
Cambridge Univ. Press (Cambridge).\\
G. Vilkovisky, \cqg {\bf 9} (1992) 895.
\bibitem{buchbinder}
I.L. Buchbinder, S.D. Odintsov, and I.L. Shapiro,
{\it Effective Action in Quantum Gravity}, IOP Publishing, Bristol 1992.
\bibitem{amendola}
L. Amendola \pl {\bf 301 B} (1993) 175.
\bibitem{asfree2}
S. Capozziello, R. de Ritis, and A.A. Marino {\it Asymptotic Freedom
Cosmology}, to appear in \pl {\bf A} 1998.
\bibitem{lambdagen}
S. Capozziello, R. de Ritis, and A.A. Marino , \grg {\bf 30} (1998) 1247.
\bibitem{any}
S. Capozziello and R. de Ritis, \ijmp {\bf D 5} (1996) 209.
\bibitem{wald}
R.M.Wald, \pr {\bf D28} (1983) 2118.
\end{thebibliography}
\end{document}